# Aboriginal Astronomical Traditions from Ooldea, South Australia
# Part 1: Nyeeruna and the Orion Story

Trevor M. Leaman and Duane W. Hamacher

Nura Gili Indigenous Programs Unit, University of New South Wales, Sydney, NSW, 2052, Australia
Emails: t.leaman@unsw.edu.au; d.hamacher@unsw.edu.au

**Abstract:** Whilst camped at Ooldea, South Australia, between 1919 and 1935, the amateur anthropologist Daisy Bates CBE (1859-1951) recorded the daily lives, lore, and oral traditions of the Aboriginal people of the Great Victoria Desert region surrounding Ooldea. Among her archived notes are stories regarding the Aboriginal astronomical traditions of this region. One story in particular, involving the stars making up the modern western constellations of Orion and Taurus, and thus referred to here as "The Orion Story", stands out for its level of detail and possible references to transient astronomical phenomena. Here, we critically analyse several important elements of "The Orion Story", including its relationship to an important secret-sacred male initiation rite. This paper is the first in a series attempting to reconstruct a more complete picture of the sky knowledge and star lore of the Aboriginal people of the Great Victoria Desert.

**Notice to Aboriginal and Torres Strait Islander people:** This paper contains brief references to Australian Aboriginal male initiation rites and its links to the sky, the full knowledge of which is not discussed here as it is restricted to senior Western Desert men. It also gives the names and images of people who are deceased.

> *"Here in the bright, still evenings, I studied the skies, astronomy being an old love of mine, and compiled my aboriginal mythologies, many of them as poetic and beautiful as are the starry mythologies of the Greeks."* – Daisy Bates (1936: 23).

## 1  INTRODUCTION

The first in-depth study of Aboriginal Australian astronomy began with William E. Stanbridge, who wrote on the ethnoastronomy of the Boorong people of western Victoria (Stanbridge, 1858; 1861). Other early pioneers in the field include Brian Maegraith (1932), Charles Mountford (1939; 1958; 1976), and Norman Tindale (1959; 2005). To this list we can add Daisy Bates CBE (1859-1951) who, whilst camped at Ooldea on the southern fringes of the Great Victoria Desert, South Australia between 1919 and 1935, recorded the language, customs, and oral traditions of the local Aboriginal people, including their astronomical knowledge and traditions.

This paper is the first of a series that comprehensively studies and analyses Aboriginal astronomical traditions in the Great Victoria Desert in western South Australia and south-eastern Western Australia. In this paper, our aim is to use data recorded by Daisy Bates to analyse one of the more detailed astronomical traditions from Ooldea, South Australia – that of "The Orion Story" (discussed in Sections 4 and 5). In this paper, we provide a brief biography of Daisy Bates and explore her astronomical interests and pursuits at Ooldea. We then briefly outline our search of the Daisy Bates Collection, held in the archives of the National Library of Australia (NLA) in Canberra. This is followed up by a more detailed analysis of an oral tradition involving the stars surrounding the constellations of Orion and Taurus, and which appears to contain references to several transient astronomical events. We then briefly look at how this story is incorporated into the male initiation rites at Ooldea, and how this may offer clues to a sophisticated understanding of the daily and annual movements of the celestial sphere.

## 2    DAISY BATES CBE (1859-1951)

Daisy Bates (Figure 1) is an enigmatic, complex, and somewhat controversial figure in Australian History. Her popular biographies (e.g. Blackburn, 1994; Hill, 1973; Salter, 1971; Wright 1979) contain a fictitious and fanciful version of Bates' early life, with claims that she was of 'aristocratic' Anglo-Irish Protestant heritage. Later investigations show that she was actually born into poverty to Irish-Catholic parents and orphaned at a young age (De Vries, 2008; Reece, 2007). Despite her poverty, she was educated in languages, literature, history, and science, all playing an important role in her later life in England and Australia (De Vries, 2008: 46-51).

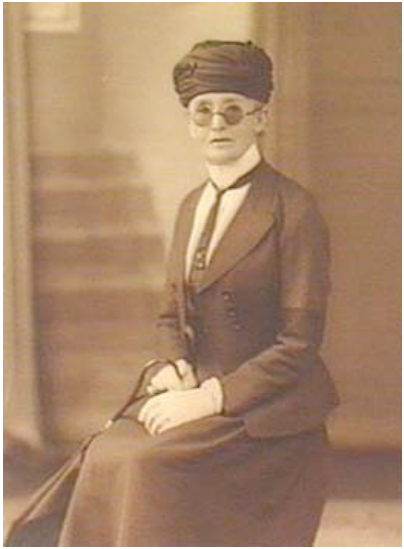

**Figure 1:** Daisy Bates in Adelaide circa 1936 (aged 76). Image credit: State Library of South Australia.

During a voyage to Australia in 1899, Bates befriended Father Dean Martelli, an elderly Catholic priest returning to an Aboriginal missionary at Beagle Bay near Broome. Over several conversations, Bates soon learned the plight of the Aboriginal people, who were then dying in large numbers from 'white man's diseases and despair', along with their culture (De Vries, 2008: 114). Showing great interest in recording and preserving their culture, Bates accepted an invitation to visit the mission, where one Abbott Nicholas was compiling a language dictionary. Apart from being Bates' first physical contact with Aboriginal people (De Vries, 2008: 115), the experience also taught her the basic skills of a field anthropologist (Reece, 2007: 36). Temporarily rejoining her husband at Roebuck Plains in 1901, she used these skills to observe and record the vocabularies and rituals of the Aboriginal people camped at their cattle station (De Vries, 2008: Chapters 9 and 10; Reece, 2007: 39; Salter, 1971).

After moving to Perth in 1904, Bates worked as a junior clerk for the Western Australian Government,

collecting and compiling vocabularies from several Aboriginal language groups. As well as enabling Bates to build up a more complete picture of Aboriginal life (De Vries, 2008: Chapter 11; Reece, 2007: Chapter 2), her work also brought her into contact with the anthropological fraternity, through which she gained some level of academic credibility, becoming the Western Australian correspondent for the Anthropological Institute of Great Britain, Fellow of the Royal Anthropological Society of Australasia, and Member of the Royal Geographical Society of Melbourne (Bartlett, 1997; De Vries, 2008: 149; Reece, 2007: 49-50).

By 1910, and after an extensive eight-month field survey of the Aboriginal peoples of Western Australia, including interviewing the last survivors of the Bibbulmun culture, Bates had amassed a huge amount of data on language, oral traditions, religion, and kinship. It was finally published posthumously in 1985 as *The Native Tribes of Western Australia*, thanks to the extensive editorial work provided by White (1985). The ethnographic information contained within it has assisted in supporting recent native title claims (Reece, 2007: 9-10; Burke, 2011; Sullivan, 1995).

After being bestowed the title of "Honorary Protector of Aborigines" in 1912 (De Vries, 2008: 165; Reece, 2007: 67), Bates spent the next 20 years among the Aboriginal people of South Australia, firstly with the Mirning people at Eucla and Yalata, then moving to Ooldea ("Yooldilya Gabbi") in 1919 (Bates, 1938: Chapters 15 and 17; Colley et al., 1989; De Vries, 2008; Reece, 2007; Figure 2).

Bates was appointed CBE (Commander of the Order of the British Empire) in 1934, more in recognition of her Aboriginal welfare work than for her anthropological research (De Vries, 2008: 215-217; Reece, 2007: 112-113). In the following year, Bates left Ooldea for Adelaide to work on her autobiography (De Vries, 2008: Chapter 18; Reece, 2007: Chapter 4). Published in 1938, *The Passing of the Aborigines* became a best-seller, praised by the general public but harshly criticised by the anthropological community, mostly for the outmoded portrayal of Aboriginal people as a "dying race" and her increasing obsession with unsubstantiated and sensationalistic stories of cannibalism (De Vries, 2008: 243-246; Reece, 2007: 124-125). In 1945, deteriorating health and poor eyesight eventually forced Bates to move to Adelaide. Bates passed away at Prospect, South Australia on 19 April 1951, aged 91 (De Vries, 2008: 262; Reece, 2007: 154-155).

## 3    ASTRONOMY AT OOLDEA

Located on the southern fringe of the Great Victoria Desert, Ooldea served as an outpost ("Ooldea Siding") for the Trans-Australian Railway (Bates, 1938: Chapters 15 and 17; Brockwell et al., 1989; Colley et al., 1989; Reece, 2007: 79). It was also the location of one of the few permanent sources of freshwater ("Ooldea Soak"), which made it an important drought refuge for many Aboriginal peoples (Tindale, 1974: 69), and an ideal starting point for several inland expeditions by colonial explorers (Brockwell et al., 1989; Gara, 1989). It also played an important role as a meeting place for Aboriginal ceremony and trade, with cultural items traded from many locations across the continent (Bates, 1938; Berndt, 1941; Brockwell et al., 1989; Colley et al., 1989).

The eastern part of the Great Victoria Desert surrounding Ooldea was the traditional lands of the Kokatha people. West of the Kokatha lands were the Ngalea lands, the principle water refuge being at Waldana Well (Bates, 1921a; Gara, 1989). Together, with other groups to the north and west, they made up part of the "Western Desert culture" or "Spinifex People" (*Pila Nguru*), sharing a similar social structure and religious beliefs, and speaking closely related dialects (Berndt, 1959: 93-95; Cane, 2002; Gara, 1989). The *Pila Nguru* of Ooldea and surrounding areas were forcibly removed to Yalata in the 1950s to make way for atomic tests (Cane, 2002).

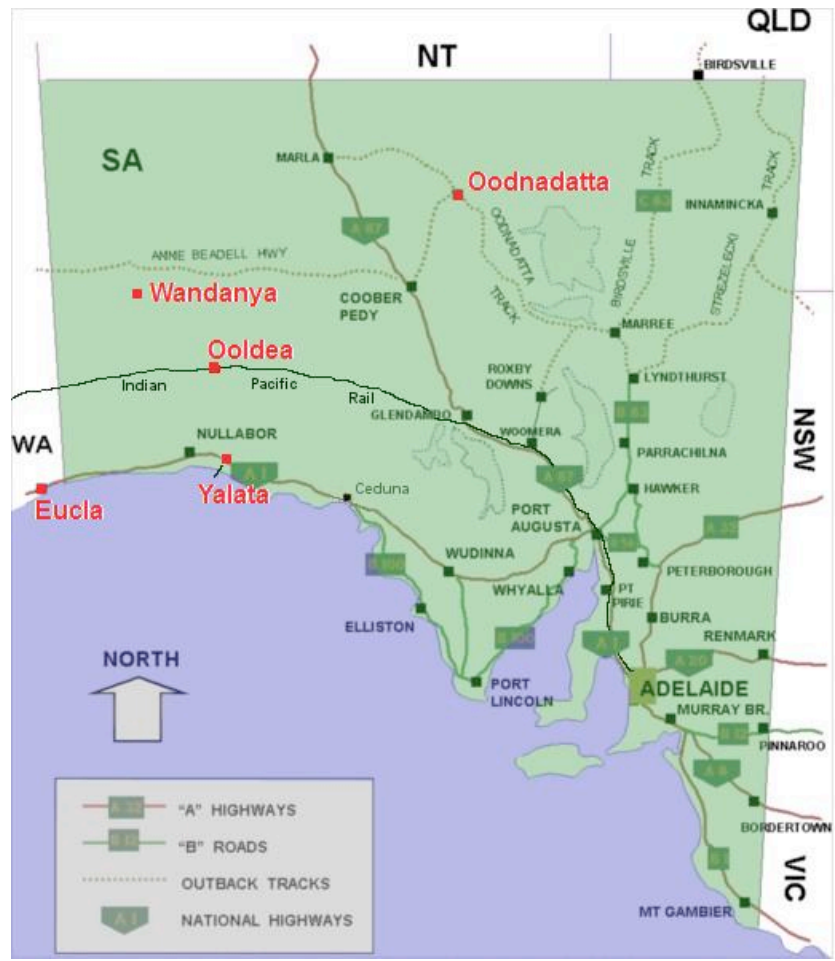

**Figure 2:** Locations where Bates lived and worked alongside the Aboriginal peoples, including Eucla, Yalata, and finally Ooldea. Also indicated are the locations of Wandanya (Waldana Well), the area from which "The Orion Story" originated according to Bates, and Oodnadatta, where anthropologists Ronald and Catherine Berndt witnessed the *Minari* and *Baba* Inma in 1944. Image credit: user Astrokey44 (Wikimedia Commons).

Whilst at Ooldea, Bates again recorded a substantial amount of material on her observations of Aboriginal daily life (Bates, 1904-1912). In time, she gained sufficient trust and respect of her Aboriginal neighbours, which not only allowed her to witness sacred ceremonies, but to be entrusted with the safe-keeping of Inma objects (sacred boards) after the death of their custodians (Bates, 1938; Reece, 2007: 95).

In her book, Bates also mentions her passion for astronomy and the manner in which she observed the Ooldea night sky (Bates, 1938: Chapter 17):

> "...a smaller bough shed on the crest of the hill, with a ladder leading to its leafy roof, that was my observatory. Here in the bright, still evenings, I studied the skies, astronomy being an old

>love of mine, and compiled my aboriginal mythologies, many of them as poetic and beautiful as are the starry mythologies of the Greeks."

Based on her archived notes, it seems her astronomical interests did not go much beyond that of an amateur. She seemed more interested in the star stories than in the astrophysical and cosmological concepts of the time. We know she did not possess a telescope at Ooldea (Reece, 2007: 95), but the use of printed star charts and ephemerides cannot be ruled out.

At Ooldea, Bates sat with elders as they drew maps of the constellations, stars, and totemic signs in the sand with sticks, which she later translated into their Western equivalents using her own knowledge of astronomy (Hill, 1973). Most of the astronomical knowledge and traditions collected by Bates was restricted to males (i.e. "Men's business"). Although the exact reason why this information was passed on to Bates is unclear, it was most likely the result of Bates' long tenure with the community, building up trust and respect with the elders, combined with the fact that she was not Aboriginal. It was not uncommon at the time for elders to share sacred or secret knowledge to outsiders provided they did not reveal it to non-initiated members of the community. Doing so could have serious repercussions, as attested by the legal case that followed the publication of Charles Mountford's (1976) book *Nomads of the Australian Desert*, which contained knowledge of some secret Pitjanjatjara ceremonies. This outcome resulted in the book being banned in the Northern Territory, and a heightened level of caution and distrust shown towards researchers since (Neate, 1982).

Some of Bates' work on Aboriginal astronomical knowledge was published, either as syndicated newspaper articles (e.g. see Bates, 1921b; 1921c; 1924a; 1924b; 1933) or in books authored by others (e.g. Ker Wilson, 1972). But much of it remains in the form of unpublished field notes in the National Library of Australia. It is from these notes that the original accounts of Aboriginal astronomy from the Great Victoria Desert, centred on Ooldea, are found. These are dispersed within Section VII of Manuscript 365 of the Daisy Bates Collection (Bates, 1904-1912) as a series of numbered folios. These folios are undated, making it difficult to place these in a chronological order. Complicating the issue is the fact that some folios appear to contain fragments of astronomical knowledge from Ooldea that were incorporated with those from other language groups or regions that Bates studied previously, such as those from the Mirning people of Eucla, or the Bibbulmun and Noongar people of southern Western Australia (Folios 25/308; 26/78; 26/81-84). Other folios (Folios 26/47; 26/106-7; 26/113) appear to be attributed solely to the Ooldea region and contain lists of astronomical objects with their corresponding Aboriginal names or traditions.

Combining these objects with contributions from other researchers (e.g. Hill, 1973; Berndt, 1941), a full list of astronomical objects, their Aboriginal name/meaning, and their Western counterpart are listed in Table 1. These data are being used for a larger project to reconstruct the Ooldea night sky. A full analysis of these astronomical traditions, including their relationship to seasonal change, food economics, and social structure, is the focus of future work.

*Table 1:* The Ooldea night sky as reconstructed from the field notes of Daisy Bates (1904-1912), with contributions from Hill (1973) and Berndt (1941). Spelling variations most likely reflect the different language groups from which the stories originated.

| Object Type | Western Name | Aboriginal Name or Attribution | Aboriginal Interpretation |
|---|---|---|---|
| Galactic | Milky Way | *Dhoogoor Yuara*[1] | River that never dries/road of dreaming |
| | Coal Sack | *Kallaia, Kalia* | Emu Head (body is dust lanes in Milky Way) |
| | Magellanic Clouds | *Boolbarradu, Balbaradu* | Brothers (collectively) |
| | Large Magellanic Cloud | *Murgaru, Badhu-Wudha* | Right-handed Brother |
| | Small Magellanic Cloud | *Oimbu, Kurulba* | Left-handed Brother |
| Constellation, Cluster, or Asterism | Crux | *Waljajinna* | The Track of Eaglehawk |
| | Pointers | *Jurding, Dhurding* | Club of Eaglehawk |
| | Delphinus | *Nyumbu, Mamu,* | Crow Children |
| | Aquarius | *Bailgu* | Brush Fence |
| | Gemini | *Wati Kutjera*[2] | Two Men (ancestral beings) |
| | Orion | *Nyeeruna, Nyiruna* | Hunter of the Seven *Mingari* Sisters |
| | Hyades | *Kambugudha* | Eldest of the *Mingari* (Thorny Devil) Sisters |
| | Pleiades | *Yugarilya, Kunggara* | Seven Young *Mingari* Sisters |
| | Pisces | *Warramula*[1] | *Kadaicha* (Sorcerer) Men on the Trail |
| | Line of stars between Beta Tauri (Elnath) and Achernar | *Mingari's Dogs* | Row of dingo puppies placed before *Nyeeruna* to stop his advances on the Seven *Mingari* Sisters (Pleiades) |
| Star | Stars (generic noun) | *Kattana* | "Heads" |
| | Alpha Centauri | *Maalu* | |
| | Beta Centauri | *Kanyala* | |
| | Alpha Geminorium | *Mumba*[2] | The lazy one of the *Wati Kutjera* |
| | Beta Geminorium | *Kuruka'di*[2] | The wise, skilful one of the *Wati Kutjera* |
| | Altair | *Kangga Ngoonji* | Crow Mother |
| | Vega | *Gibbera* | Bush Turkey |
| | Antares | *Warrooboordina*[1] | Black Cockatoo (Fire Carrier) |
| | Rigel | ? | *Badwuja's* Brother (?) |
| | Canopus | *Joor-Joor, Jurr-jurr* | The Owlet Nightjar |
| | Beta Tauri (Elnath) or Zeta Tauri (?) | *Babba* | Dingo Father |
| | Aldebaran | | Left Foot of *Kambugudha* |
| | Betelgeuse | | *Nyeeruna's* Right Arm |
| | Achernar | *Ngurunya (?)* | Dingo Mother |
| | Spica | *Karduna* | ? |
| Solar System | Morning Star | *Maalu*[3] | Red Kangaroo |
| | Evening Star | *Kulbir*[3] | Grey Kangaroo |
| | Venus | *Genba (Guldu) Katta* | *Genba's* (*Guldu's*) Head |
| | Mars | *Kogolongo, Koggalangu* | Black Cockatoo with red feather in its tail |
| | Jupiter | *Karrail Katta* | *Karrail's* Head |
| | Moon | *Beera* | *Beera Goarrija* (Waxing), *Beera Bulgana* (Full), *Beera ilung* (Waning) |
| | Lunar Eclipse | *Beera Dharbongu* | |
| | Meteor | *Mama* | |

[1] From Hill (1973) and may or may not refer to that same group. Other star names are similar. [2] From Berndt (1941). The *Wati Kudjera* story originated from the Warberton Range, Western Australia, but had drifted down to Ooldea through tribal migration. These stars and story are not mentioned in any detail in Bates' notes. [3] Bates incorrectly identifies and describes Jupiter as the morning star and Venus as the evening star, when in fact both "stars" are Venus.

**4      "THE ORION STORY"**

"The Orion Story" stands out for its detail and its intriguing references to possible observations of several transient astronomical events. This story is first encountered in an article published in The Australasian (Bates, 1921b). Whilst giving an account of bird life in the region around Ooldea, she digresses into the story:

> "...*Jurr-jurr*, a species of night owl[1], whose hoarse cry is thus rendered by the natives, has a distinction of being translated into a star, and is now Canopus, watching over *ming-arri* (mountain devil), now the Pleiades, who is being pursued by *Nyiruna* (Orion) 'round and round the sky'. The little mountain devil, inaptly named (as it is absolutely the most harmless of all living creatures) occupies a unique position in native legend. *Ming-arri* were all women in long ago times who never wished to mate with men. They lived by themselves and kept a tribe of dingoes to keep all men away, the dogs killing and eating all the men they caught. *Ming-arri* brought forth and reared their babies, but laid the injunction on each one as it grew up that "it must never talk or whistle" or the men would catch it. *Nyiruna* was a great hunter in those days, and he wanted *ming-arri* very badly for his wives, and left food for them and tried to catch them, but the dingoes ate the food and chased *Nyiruna* away, and by and by, when *ming-arri* went into the sky, *Nyiruna* followed them, and there he is, still chasing them round and round, while the dogs, who are all around *ming-arri*, still keep him away. *Ming-arri* have now no voice at all, because their mothers never let them speak in the old days."

This article seemed to have attracted the interest of one of her readers as the next time the story is mentioned by Bates it is in the form of a response to "Canopus" in the editorial section of *The Australasian* (Bates, 1904-1912: Folio 25/441-442; Bates, 1921c):

> "The myth referred to, in my article on "The Great Plain's Edge" does not belong to the Nullarbor Plain natives, for there are no natives living on the plain except a few "strays" from Eucla or the north, who work now and then at White Well and Nullarbor Plains stations, near the head of the Bight. The myth concerning Orion and the Pleiades belongs to a tribe living near Wandunya and other waters, about 128 miles north 200 miles northwest of Ooldea Siding. The myth is known to the *Mingarri* totem people, to whom it belongs, so to speak; and as the myth is a totem one, and its totemists have only recently come into civilised areas, your correspondent "Canopus" may rest assured that it is purely an aboriginal myth, with no "dressing" whatever from outside. The country from which the *Mingarri* totem people come is not yet taken up, and those who have come to my camp from that area are absolutely uncivilised. Up to the present I have obtained only a portion of the myth, but with every fresh arrival I obtain a little more. No new narrator has contradicted the portion of the myth obtained. Orion is *Nyiruna* chasing the Pleiades, but I am not yet able to name the particular stars which are *Mingarri's* dogs, though Achernar has been pointed to more than once, as one of *Mingarri's* dogs. The natives' personal description of Orion and the Pleiades is not suitable reading for other than purely scientific magazines, but it is extremely interesting and quite "native". I have not the whole connected myth as yet, for it takes a long time to get a complete legend or myth from the uncivilised natives. I may mention that in all native star myths, from the Kimberley district in Northern Western

> Australia to the *Mingarri* totem group in South Australia, the Pleiades are "a lot of women"; but as far as I remember without my notes it is only in the group that Orion comes in as *Nyiruna* hunting them. The *Mingarri* myth is known to neighboring tribes of other totems."

The oral tradition in question is a story of the *Mingarri* totem from an Aboriginal community near Wandunya (also known as Waldana Well, ~322 km NW of Ooldea, Figure 2). According to Bates, this oral tradition is known from a wide area of Central Australia, beyond the border into Western Australia to Diamantina River and Cooper River regions of northeastern South Australia - southwestern Queensland to the eastern edge of the Great Nullarbor Plain of western South Australia. *Mingarri* is the totem of Thorny Lizard (*Moloch horridus*), also known as the thorny dragon, thorny devil or mountain devil (Figure 3). This small lizard (up to 20 cm in length) is covered in conical spikes and inhabits the desert and scrub over most of central Australia (Browne-Cooper et al., 2007).

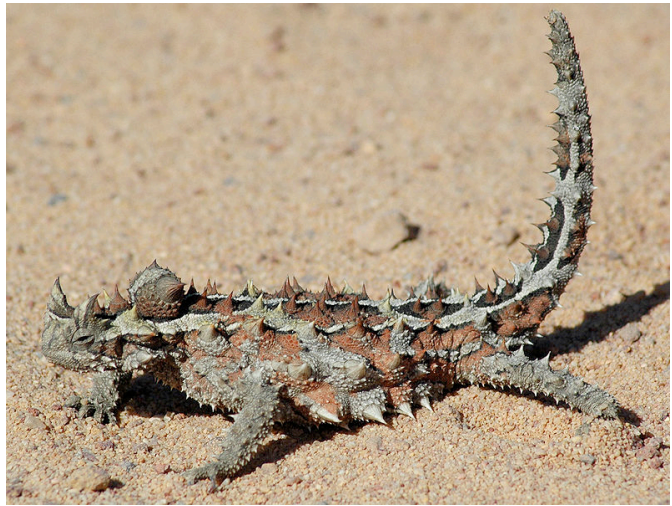

**Figure 3:** According to Bates' story, the Seven *Yugarilya* Sisters of the Pleiades were all *Mingarri* (Thorny Devil, or lizard) totem. The Thorny Devil (*Moloch horridus*) is found throughout the central desert regions of Australia and feeds exclusively on ants (*Minga*). The "hump" on the back of its neck is the "false-head", used as a defense mechanism. Image credit: user KeresH (Wikimedia Commons).

From the tone of Bates' response we can infer that her correspondent "Canopus" has raised doubts to the genuineness of the story, possibly as it resembles the Greek version of the Orion myth too closely to be purely "native". Bates goes to great lengths to assure that the story is genuine but fragmentary, and that she is still in the process of collecting more parts of the complete story from different sources. This may explain the twelve-year gap between her first, incomplete version of "The Orion Story" (above) and the final, more complete version that she later published in The Sydney Morning Herald (Bates, 1933). This version can also be found in Bates' manuscript records as Folio 25/85-88. A second, almost identical folio (Folio 26/13-16) also exists in the records but it cannot be dated in relation to the other. Their similarity suggests they were both written about the same time, and one may have been a "back-up copy" for the other. As both are similar, only one is reproduced in full in the Appendix.

In these accounts, the stars that constitute the Western constellation of Orion (Figure 4) are seen as a hunter, named *Nyeeruna* (spelled *Niyruna* in Bates' earlier account)[2]. He is a vain pursuer of women, with a feathered headdress, ochred body, string belt (Belt of Orion), and whitened tassel (the scabbard of Orion's sword). Each night he pursues the sisters of the Pleiades (*Yugarilya*), who are of the *Mingari* (also spelled *Min-garri*, *Ming-arri* or *Mingarri*) totem.

*Nyeeruna* is forever prevented from reaching *Yugarilya* by *Kambugudha*, their eldest sister, represented by the Hyades, who guards her younger sisters. *Kambugudha* taunts *Nyeeruna* by standing naked before him with her legs spread (represented by the V-shape of bright stars in the Hyades, the "head" of Taurus the bull). The club in *Nyeeruna's* right hand (Betelgeuse) fills with "fire magic" ready to throw at *Kambugudha*. However, she defensively lifts her left foot (Aldebaran), which is also full of "fire magic", which causes him great humiliation and puts out the fire magic of his arm. In her contempt of his vanity, *Kambugudha* places a line of dingo puppies[3] between her and *Nyeeruna*, represented by an arc of stars between Orion and the Hyades.[4]

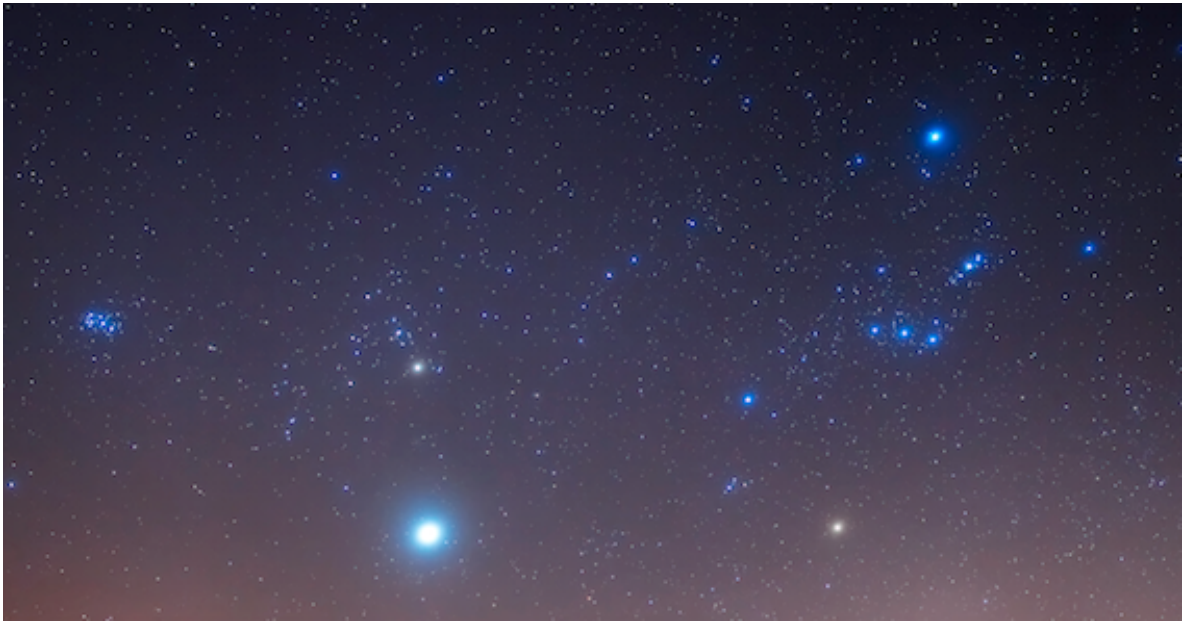

**Figure 4:** A correct Southern Hemisphere perspective of the night sky containing most of the key players in "The Orion Story", including (from left to right) the Pleiades (*Yugarilya*, the seven *Mingari* Sisters), The Hyades (*Kambugudha's* legs), The "horns of the bull" (*Babba* the father dingo), Orion's "shield" (Dingo puppies) and Orion (*Nyeeruna*). The bright object below the Hyades is the planet Jupiter. (Image credit: Free stock image from www.favewalls.com).

Eventually, *Nyeeruna's* magic returns with force and his hand (Betelgeuse) increases in brightness. *Kambugudha* calls to *Babba* the father dingo, who rushes over to *Nyeeruna* and "shakes and swings him east and west by his middle" while *Kambugudha* points and laughs at him. However, her timid sisters are frightened and hide their heads until *Babba* loosens his hold and returns to his place. As this happens, many other beings such as *Joorrjoorr* the owlet-nightjar (Canopus), *Beera* (Moon), and *Kara* the red-back spider (Rigel), mock and laugh at *Nyeeruna*, who again loses his red fire and "no sparks" come from his body in his shame and humiliation.

Mountford (1948: 167-168) records a variation of the *Nyeeruna* story: the stars of Orion are *Nirunya*, a man pursuing the group of women called *Kunkarunkara* (Pleiades). Mountford does not give many details, but claims that the women usually outsmarted Nirunya. On occasion, one of the women "would fall victim to his desires" (p. 167). The Mountford version (which came from an Aboriginal woman named Numidi), from the Central Desert, describes a place where Nirunya attempted to capture two women as they were digging for yams. The women saw him coming from the sky and went underground briefly before bursting out and escaping to the sky. Physical traces of the incident are evident on the ground, including the place from which the women emerged and a hole dug in the rock where they were searching for yams. Unfortunately, the Mountford version does not provide any details of other stars that may shed light on the story recorded by Bates. Maegraith (1932), who wrote about the astronomy of the Aranda (Arrernte) and Luritja peoples of the Central Desert, did not collect stories about Orion or the Pleiades, as they were not visible in the early August night during his fieldwork in Hermannsburg (Ntaria). Tindale (1959) recorded a story from the Western Desert that describes the Pleiades as Kungkarungkara and Orion (stars of the belt) as Njiru.

## 5    INTERPRETING "THE ORION STORY"

Of the various Aboriginal traditions across Australia regarding the stars in Orion and the Pleiades, nearly 90% associate the stars of Orion with a man or group of men and the stars of the Pleiades with a woman or group of women – a trend found across the world (Fredrick, 2008: 57). Although there are similarities between the Greek myth of Orion and Bates' record of the Orion Story, there is no evidence of post-colonial Western cultural influence. In fact, the story forms the basis of an important male initiation rite (Berndt and Berndt, 1943; 1945, see Section 6). Under closer scrutiny, the story unveils several very interesting elements.

Firstly, the description of *Nyeeruna's* arm (Betelgeuse) is that it fills with fire magic, his hand becoming periodically brighter, and then fainter before brightening again. This suggests, as first proposed by Fredrick (2008: 59), that the Aboriginal observers may have noticed the variability in brightness of the star Betelgeuse. Betelgeuse is a semi-regular variable star with a period of ~400 days (Dupree et al., 1987; 1990; Gray, 2000; 2008; Kiss et al., 2006; Smith et al., 1989; Stothers, 2010). Although the magnitude range, from maximum to minimum brightness and back again, is easily noticeable by eye ($m_{v(max)}$ = 0.1, $m_{v(min)}$ = 1.1, $\Delta m_v$ ~1.0, see Table 2), Betelgeuse would need to be observed over many cycles spanning several years for its variable nature to be noticed. However, such a feat is not outside the realms of possibility for keen Aboriginal observers. The close proximity of Rigel and Aldebaran to Betelgeuse enables both to be used as reference stars, which aid in determining Betelgeuse's brightness excursions visually. This technique is still employed by modern variable star observers (e.g. Sigismondi, 2000) and was the method employed by Herschel that lead to his discovery of Betelgeuse's variability in 1836 (Herschel, 1840a; 1840b). Interestingly, Bates indicates that this increase in "fire and lust" may be due to the effects of "radiations from nebulae" (Appendix) suggesting that she is unaware of the variable nature of Betelgeuse and its possible connection to the story.

Secondly, *Kambugudha's* foot (Aldebaran), like *Nyeeruna's* right hand (Betelgeuse) also fills with "fire-magic", suggesting that it, too, was observed to be variable. While Aldebaran is indeed a small-amplitude variable star (Henry et al., 2000; Wasatonic and Guinan, 1997), the brightness variations are much too small to be noticed by naked eye observers ($m_{v(max)}$ = 0.85, $m_{v(min)}$ = 0.88, $\Delta m_v$ ~0.03). This suggests that the "fire magic" description may not relate to observed stellar variability. Instead it may refer to the intrinsic reddish-orange (i.e. "fire-like") colour of both stars, and the effects of atmospheric scintillation at low elevations. Alternatively, it is reasonably plausible that Betelgeuse's intrinsic brightness variations *were* indeed noticed, but the same 'qualities' were also bestowed on Aldebaran by the storytellers to add a sense of drama to this part of the story. Without further substantiating evidence we cannot prove either hypothesis. We are currently searching the anthropological records for other

Indigenous references to the observed variability in Betelgeuse.

*Table 2*: Bates' original attribution of astronomical characters to principle stars in the Aboriginal story of *Nyeeruna* and *Kambugudha*, listing their Aboriginal and Western name, Bayer designation, spectral type, visual magnitude ($m_v$), variability (Yes or No), their magnitude range ($\Delta m_v$). Of the stars described in this oral tradition, only the variability of Betelgeuse would be noticeable to the naked eye.

| Aboriginal | Western | Designation | Spectral | $m_v$ | var | $\Delta m_v$ |
|---|---|---|---|---|---|---|
| *Yugarilya* (seven *Mingari* Sisters) | Pleiades | M45 | ... | ... | ... | ... |
| Left Foot of *Kambugudha* | Aldebaran | α Tauri | K5 III | 0.87 | Y | 0.03 |
| *Nyeeruna's* Right Arm | Betelgeuse | α Orionis | M2 Iab | 0.60 | Y | ~1.0 |
| *Joorrjoorr* the Owlet Nightjar | Canopus | α Carinae | F0 1b-II | -0.74 | N | ... |
| *Kara* the Redback Spider | Rigel | β Orionis | B8 Ia | 0.13 | Y | 0.05 |
| *Babba* the Father Dingo | Elnath? Zeta Tauri? | β Tauri ζ Tauri | B7 III B4 III ep | 1.68 2.99 | N Y | ... 0.10 |
| Mother Dingo | Achernar | α Eridani | B3 IV ep | 0.45 | Y | 0.03 |
| ? | Procyon | α Canis Minor | F5 IV | 0.38 | N | ... |

Thirdly, the story contains a reference to *Nyeeruna* not having "sparks" issuing from his body after being humiliated by *Kambugudha*, which, by inference, suggests that "sparks" may issue occasionally, perhaps when he is "filled with lust" for the seven *Mingari* sisters. Bates again attributes this phenomenon to "nebulae" (Appendix). The "sparks" are a possible reference to the nearby Orionid meteor shower, caused by the earth passing through the dust stream of Comet Halley. The radiant of the Orionids is very close to Betelgeuse and Orion's "club" (Figure 5) and typically peaks over the last two weeks of October each year (McIntosh and Hajduk, 1983). During this time, Orion rises around midnight and is high in the sky before dawn. Peak intensities can vary from year to year due to clumping of meteoroid material in orbital resonant regions (Rendtel, 2007; Štohl and Porubčan, 1981; Trigo-Rodrigez et al., 2007) and large showers have been recorded historically from many cultures (Ahn, 2003). Meteors feature prominently in Aboriginal traditions (Hamacher and Norris, 2010) and are generally given negative associations, including portents of death and punishment for breaking laws and traditions (Hamacher, 2011; Hamacher and Norris, 2010).

Next, the attribution of *Kara* the red-back spider with the blue star Rigel (β Orionis, spectral type B8 Ia, $m_v \sim 0.13$) is puzzling. In fact, Bates describes Rigel as "redly shining" (Appendix) revealing complications in Bates' account, and leaving one to question the reliability of the story as a whole. After further investigation of the archival documents, two other folios were uncovered (Folios 26/78 and 26/81-82, dates unknown), both containing a haphazard list of Aboriginal names and their attributions to stars, asterisms, and constellations. It is possible that these represent Bates' earliest attempts at piecing together Aboriginal astronomical traditions from the small fragments of information given to her from her Aboriginal informants. Near the end of Folio 26/78, titled "Mythical Names of Stars", the following passage appears:

> "…*Ngurunya* is a star which sets at 9pm in March (Achernar). *Kara* (spider) is northeast and is the winter evening star. He comes close up to *Ngurainya* (Vega)."

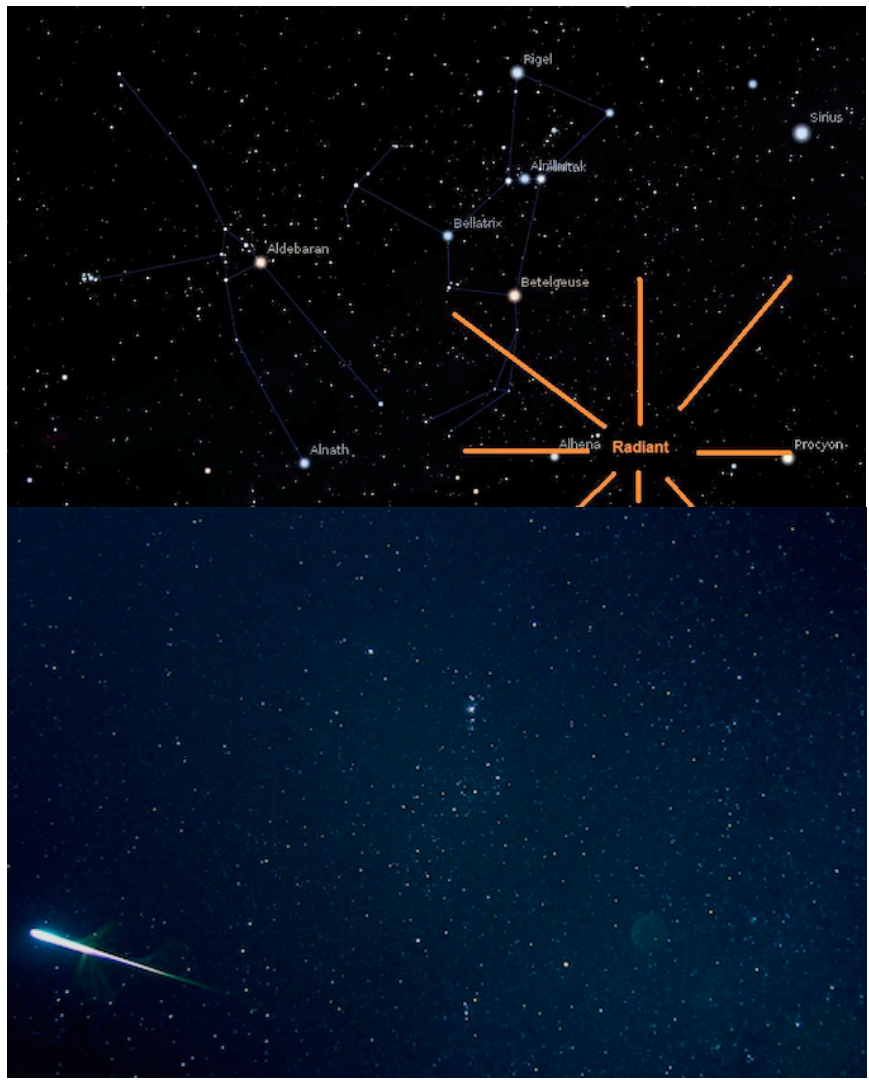

**Figure 5:** (a) The radiant of the Orionids in relation to Betelgeuse in Orion (*Nyeeruna*) and Aldebaran in Taurus (*Kambugudha*). (b) A "spark" from *Nyeeruna's* arm shooting across to *Kambugudha* (an Orionid caught mid-flight). Note: both images have a southern hemisphere perspective. (Image credits: (a) image generated using Stellarium (www.stellarium.org). (b) Photograph by Rich Swanson, Sierra Vista, Arizona).

At Ooldea, the star Achernar sets at ~ 21:00 in mid-May, not in March as Bates indicates. In March, Achernar sets at ~ 02:00. This discrepancy could be blamed on inaccurate timekeeping (did Bates use a timepiece, or did she just estimate the time of day/night?) or the result of a simple typographical error.

The rest of the passage seems to identify an alternative candidate star for *Kara*. The star that best matches her statements "star in the northwest" and "winter evening star" is Arcturus (α Boötis), which rises before Vega. The phrase "He comes close up to *Ngurainya*" is most likely a verbatim description given to Bates from her informant/s. The attribution of *Kara* with Arcturus is a much better match than for Rigel, and solves the ambiguity in Bates' recording of the Orion Story. Arcturus is both conspicuously bright ($m_v = -0.04$), and of the right colour (spectral type K1.5 IIIp, making it appear distinctly orange). Arcturus also plays an important role in the astronomical traditions of other

Aboriginal groups throughout Australia[5] thus strengthening its candidacy here. Table 2 has been amended accordingly to incorporate this new information (see Table 3).

*Table 3*: Same as Table 2, but with re-attribution of Kara the Redback spider with Arcturus, an orange-red giant star, based on Bates' unpublished notes contained in Folio 26/78. The same Folio also gives a possible name to the Mother Dingo. Procyon's Aboriginal name and role still remains unclear.

| Aboriginal | Western | Designation | Spectral | $m_v$ | var | $\Delta m_v$ |
|---|---|---|---|---|---|---|
| *Yugarilya* (seven *Mingari* Sisters) | Pleiades | M45 | ... | ... | ... | ... |
| Left Foot of *Kambugudha* | Aldebaran | α Tauri | K5 III | 0.87 | Y | 0.03 |
| *Nyeeruna's* Right Arm | Betelgeuse | α Orionis | M2 Iab | 0.60 | Y | ~1.0 |
| *Joorrjoorr* the Owlet Nightjar | Canopus | α Carinae | F0 1b-II | -0.74 | N | ... |
| *Kara* the Redback Spider | Arcturus | α Boötis | K1.5 IIIpe | -0.04 | N | ... |
| *Babba* the Father Dingo | Elnath? <br> Zeta Tauri? | β Tauri <br> ζ Tauri | B7 III <br> B4 III ep | 1.68 <br> 2.99 | N <br> Y | ... <br> 0.10 |
| *Ngurunya* (?) the Mother Dingo | Achernar | α Eridani | B3 IV ep | 0.45 | Y | 0.03 |
| ? | Procyon | α Canis Minor | F5 IV | 0.38 | N | ... |

But why the error? According to De Vries (2008: 168-169), in the years since arriving at Ooldea, Bates was gradually succumbing to the condition known as vascular dementia, most likely brought on from many years of poor nutrition and advancing age. Vascular dementia is a debilitating disease resulting in gradual memory loss and other cognitive dysfunctions (Tomimoto, 2011). It is possible that during the process of re-writing "The Orion Story" for her newspaper article, she was unable to recall all the facts of the story and, misreading her notes, inadvertently substituted Rigel for Arcturus, not realising the colour disparity between stars. Lending some weight to this hypothesis is a small passage from Folio 26/81-82 (undated), which reads:

> "*Kara* (spider) was 'mate' for *M'maingurru* (Orion). (He is opposite Bijil.)."

This appears to be an earlier fragment of "The Orion Story", and perhaps contemporary with Bates' first article. If we assume that "Bijil" is a misspelling of "Rigel", and "*M'maingurru*" is another phonetic variant of "*Mingari*" (and therefore relating to the Pleiades, not Orion), then we can see where the substitution may have taken place. In her confused state of mind, Bates may have transposed these words to read:

> "*Kara* was mate for *Mingari* (opposite Orion). (He is Rigel)."

The first statement is most likely correct[6], but the second is in error. Also, the original statement "He is opposite Bijil (Rigel)" is ambiguous. Two possible interpretations are that it either means *Kara* is opposite Rigel in Orion, making it Betelgeuse (right colour, but wrong attribution, as the star is already identified as the right arm of *Nyeeruna*) or that it means *Kara* is opposite Rigel in the sky, which makes sense if *Kara* is Arcturus, as it starts to rise in the eastern sky as Rigel sets in the west.

The only linguistic link we can find between *Kara* and spider is in the Noongar language of southwest Western Australia[7], a language probably familiar to Bates. This suggests that one of her informants at Ooldea was either originally from that part of Australia, or was at least familiar with that language. Reed (1993: 127-29) mentions a story of the "Spider Woman of the Great Victoria Desert", who amorously pursues and captures a young non-initiated boy, and takes him into the sky where they both become stars (c.f. *Kungkapanpa*, see Note 6). Although the woman is unnamed, only mentioning that she is of the "Spider Clan" (i.e. is totemically linked to spiders), there is every possibility that this story is the basis behind Bates' account of *Kara* the Redback Spider in "The Orion Story". We are investigating this aspect further.

Lastly, according to Bates, *Babba* the Father Dingo plays an important role in the story. Apart from mentioning that he is associated with the "horn of the bull", she does not actually name or indicate a particular star. Two possible candidates are Elnath (β Tauri, $m_v$ = 1.68) or the less prominent Zeta (ζ) Tauri ($m_v$ = 2.99), both stars marking the tips of the horns of Taurus the bull (Tables 2 and 3). A more intriguing possibility is that it may also relate to an eyewitness account of SN 1054, a bright supernova that was prominent in this part of the sky in the year 1054 CE (Collins et.al., 1999; Mayall, 1939; Polcaro and Martocchia, 2006). The description of *Babba* "rushing over to *Nyeeruna*" and "returning to his place" (Appendix) could be in reference to the brightening and dimming of the supernova. This may be explored in later research, though Hamacher (2014 – this volume) demonstrates the extreme difficulty in linking Indigenous astronomical traditions with historical supernovae.

## 6    "THE ORION STORY" AND MALE INITIATON RITES

In the closing paragraphs of "The Orion Story" (Appendix), Bates makes reference to witnessing a re-enactment of the story in ceremony by Aboriginal men of the Ooldea region. Although short on detail and punctuated with some personal bias, there is enough information to suggest that Bates is in fact witnessing the *Minari & Baba Inma* ("*Inma*" being the word for "ceremonial ritual" or "ceremonial paraphernalia" among this Aboriginal language group) that was later observed and recorded by anthropologists Ronald and Catherine Berndt at Ooldea (Berndt & Berndt, 1943) and Macumba Station near Oodnadatta, South Australia (Berndt & Berndt, 1945; Figure 2). This ceremony involves male elders enacting the roles of *Nyeeruna* (phonetically spelled *Nji:rana* by the Berndts), the seven *Mingari* (*Minari*) sisters, and *Babba* (*Baba*) the Dingo Father, who attacks and dismembers *Nyeeruna*. The ritual concludes with the subincision of new initiates, signifying their entry into manhood (Berndt & Berndt, 1943). The subincision itself most likely represents the act of *Nyeeruna's* dismemberment by *Babba*, and therefore by inference the initiate "becomes" *Nyeeruna* (Berndt & Berndt, 1943; 1945; 1977).

Elements of the extended ceremony observed at Macumba Station were performed day and night over the week of 11-17 June 1944, and coincided with a New Moon on the evening of the 11 June (Berndt and Berndt, 1945: 239-240). The subincision rite was performed sometime between sunrise and midday on the last day (*ibid*: 249-50).

Noting the strong link between this important ceremony and the constellation of Orion, this timing is interesting for two reasons. Firstly, due to the close proximity of the Sun to Orion at this time of year (the ecliptic runs close to Orion's "club"), Orion would not be visible in the sky at any time of the night, including at sunrise or sunset. And secondly, the timing of the rite coincides with Orion being above the horizon in the daytime sky. Whether this is intentional or coincidental cannot be determined as there are no other dated records of this ceremony at hand for comparison. If intentional, as hinted in Bates' account (Appendix), it suggests a sophisticated level of understanding of the daily and annual movements of the celestial sphere, and good positional awareness of important stars and constellations, including those unseen in the sky during daylight hours. The reason for this timing could be purely esoteric; elders possessing secret sky knowledge may know when the unseen Orion (*Nyeeruna*) was above the horizon, where this cultural hero could secretly "look down" on and "participate" in the rite, whereas new initiates lacking this knowledge are unaware and oblivious of this fact until this sky knowledge is passed on.

## 7    SUMMARY

This is the first paper in a series analysing Aboriginal astronomical traditions in the Great Victoria Desert. Here, we analysed several elements making up "The Orion Story", the most detailed of Bates' stories of the Ooldea night sky. A summary of our analysis concludes that:

The waxing and waning "fire magic" of *Nyeeruna's* (Orion's) right arm is suggestive of the observed variability in Betelgeuse. However the fact that Aldebaran is also described in these terms makes this interpretation difficult without further supporting evidence. The alternative hypothesis is that it relates to the observed effects of atmospheric scintillation at low elevations.

The "sparks" being issued from *Nyeeruna* in his lust for the seven *Mingari* sisters (Pleiades) is most likely based on observations of meteors from the Orionids, the radiant of which is close to the right arm (Betelgeuse) and "club" of Orion. The fact that the Orionids peak in mid- to late-October, when Orion is low on the Eastern horizon for most of the night prior to sunrise, also lends some weight to the "fire magic"- atmospheric scintillation hypothesis mentioned above.

The relationship between *Babba* the father Dingo and the "horn of the bull" requires further analysis. Although we offer two possible candidate stars, Elnath (β Tauri) and zeta (ζ) Tauri, the fact that Bates does not actually name either star in her story leaves this open to interpretation. One possible reason for this is that she may not have known the name of the star being pointed out to her by her informant/s, only knowing its relationship to the rest of the constellation of Taurus. Because of the location we offer a third alternative, that *Babba* may have been the bright supernova of 1054 CE. However, without substantiating evidence this hypothesis remains speculative, and we are searching the literature for other references to this event.

The orange star Arcturus better matches the colour description and position of *Kara* the Redback Spider, as given in Folios 26/78 and 26/81-82, than the blue-white star Rigel, mentioned in Bates' original account of the story (Appendix). We suggest that this ambiguity in the story may have been due to Bates' poor health and mental state at the time "The Orion Story" was transcribed.

Based on a detailed account of the *Minari* and *Baba Inma* recorded near Oodnadatta, we suggest that this male initiation rite was being timed to coincide with the few days of the year when, due to the Sun's proximity to Orion (*Nyeeruna*), it is unseen throughout the night, but is always in the sky during the daytime. If this is the case, it demonstrates a good working knowledge of the annual and daily movements of the celestial sphere and positional awareness of stars and constellations in the daytime sky. We are currently looking for further supporting evidence of this.

## 8    NOTES

1. Although Bates identifies the bird as a "night owl" in her earlier news article (Bates, 1921b), she later identifies it as an "owlet nightjar", a bird totally unrelated to owls. This is most likely the Australian Owlet Nightjar (*Aegotheles cristatus*), found throughout the Australian Outback and known for its nocturnal call.

2. Spelling variations between accounts may be due to the slightly different pronunciations of the names by informants from different language groups and/or an attempt by Bates to get the phonetics right.

3. Among many Central Desert Aboriginal communities, dingoes were domesticated and used for warmth at night, while dingo pups were used as both pets and a food source.

4. Bates describes these as being a line of stars stretching from the horns of Taurus to Achernar. Based on this description, these are most likely the 'shield' stars, $\pi^{1,2,3,4,5,6}$ Orionis and $o^{1,2}$ Orionis, and possibly some stars in Eridanus, e.g Cursa (β Eridani).

5. For instance, the Boorong of northwest Victoria call Arcturus *Marpeankurrk*, a wise woman who showed her people how to harvest the edible larvae (*bittur*) from ants' nests (Stanbridge, 1858; 1861). The acronychal rising of Arcturus in August/September marks the time to harvest the *bittir*.

6. In the Pitjantjatjara language (Tindale, 1959), the Seven Sisters are the *Kungkarungkara* (*Kunga* = young woman). This linguistically links the *Mingari* sisters to *Kara*. Interestingly, the word *Kara* is the Pitjantjatjara name for Curly Wire Grass (*Aristida contorta*). Similarly, the Anangu version of the Seven Sisters (*Kungkurangkalpa*) contains the word *Kalpa*, which may refer to Rat's Tail (*Dysphania kalpari*), an herb whose seeds are ground up and mixed with honey (from honey ants, *Tjala*) to make cakes. Phillip Clarke (pers. comm.) suggests that a more likely derivation is from the Yankunytjatjara word *Kungkapanpa*, a "female cannibal" or "bogey-woman" that steals babies and children (Goddard, 1996: 42; Goddard and Wierzbicka, 1994: 232).

7. South West Aboriginal Land & Sea Council website. URL: http://www.noongarculture.org.au/

# 9 ACKNOWLEDGEMENTS

We would like to dedicate this paper to the current descendants and past ancestors of all Aboriginal Australians, for they are true astronomy pioneers. We thank Lea Gardam (South Australian Museum) and Tom Gara for assistance with the Daisy Bates archives, and Dr David Frew (Macquarie University) for information about Betelgeuse's variability. We would also like to thank Dr Phillip Clarke (Griffith University) and Emeritus Professor Ralph Rowlett (University of Missouri) for providing useful criticisms and feedback on the manuscript. This paper began as a major project assignment for Trevor Leaman whilst enrolled as a Master of Science student at Swinburne University of Technology, but completed whilst he was a PhD candidate at the University of New South Wales. Elements of this paper were presented at the 2013 Australian Space Sciences Conference at the University of New South Wales.

## 11    APPENDIX

References to possible observed transient astronomical events discussed in this paper can be found in two folios (Folios 25/85-88 and 26/13-16) within the archives of the Daisy Bates collection (Bates, 1904-1912) at the National Library of Australia (NLA). As one appears to be a duplicate of the other, only one of these (Folio 26/13-16) is reproduced here, verbatim and in full:

Folio 26/13-16: Central Australian Astronomy: The Constellation Orion ("The Orion Story"):

*The constellation Orion is known to the Central Australian natives as Nyeeruna, a name which would seem to have some linguistic affinity with Orion.*

*Nyeeruna is a hunter, but of women only, a baffled and humiliated hunter, kept for ever at bay by Kambugudha (the "V" in Taurus bull's head), the elder sister of' Yugarilya, the Pleiades, whom Nyeeruna is ever trying to capture and possess, but they are so well-guarded by their elder sister that Nyeeruna has never been able to reach them.*

*Kambugudha always stands naked before him, feet and legs wide apart, her left foot (Aldebaran) filled with fire magic, which She threateningly-lifts each time she sees Nyeeruna's right hand (Betelgeuse) endeavoring to put red fire magic into his club, to hurl at her and so gain possession of her younger sisters. Kambugudha dares Nyeeruna with her whole body, and is so contemptuous of him and his vain personal display of feathered headdress and ochred body, string belt and whitened tassel that she has placed a line of puppies only between her and Nyeeruna (a faint waving line of stars between Orion and V in Taurus).*

*The puppies' fathers and mothers - all relations of Kambugudha - and her young sisters stand apart on roundabout tracks watching the game. The younger sisters (Pleiades) are very timid and when they see*

*Nyeeruna's body reddened with fire and lust (radiations from nebulae?), fear comes upon them and they change into Mingari (Moloch horridus [the "Thorny Devil" lizard]) while rage lasts; but Kambugudha never changes her defiant attitude and she too can emit fire from her body, so that the red fire of her anger and her magic is so strong that it can subdue the fire magic Nyeeruna throws out, and when she advances towards him, lifting her left foot, she frightens him so greatly that the fire magic of his arm becomes faint and dies out for a while.*

*Again Nyeeruna's magic comes back in great force and brightness, and when Kambugudha sees the strong magic in arm and body, she calls to a father dingo (horn of the Bull) to come and humiliate Nyeeruna and Babba the Dingo rushes over to Nyeeruna and shakes and swings him east and west by his middle and Kambugudha points at him and laughs but her frightened little sisters hide their heads under their little mountain devil neck humps until Babba loosens his hold and returns to his place again.*

*A great portion of the constellations and stars Surrounding Orion form part of this great Central Australian myth, Procyon, Achernar, Taurus and others are all ready to help Kambugudha. They resent Nyeeruna's humiliating position and they laugh and are friendly with Kambugudha because of her care for her younger sisters, the Pleiades.*

*Even Joorrjoorr (Canopus) the owlet-nightjar, though only an onlooker, laughs his Joorrjoorr laugh as he watches Kambugudha blazoning all her charms before the baffled Nyeeruna, daring him forever. Kara the red back spider (Rigel) is also redly shining, ready to bite Nyeeruna. All the animals and birds round and about jeer loudly when they see Babba the Dingo debasing Nyeeruna's manhood. Beera the moon also mocks at him whenever he sits down beside Kambugudha and her young sisters during his journeys to the west, and Nyeeruna loses his red fire and no sparks come from his body (nebulae) in his shame and humiliation.*

*On fine bright starlight nights, the old men of the Central groups watch the game between Kambugudha and Nyeeruna; the little line of star puppies between them brightens and laughs, and Achernar, the mother dingo, standing at the end of her long row of puppies, joins in the laugh, and the old men re-tell old story, and wink at Beera the moon whenever they see him beside Kambugudha and her young sisters and leering and jeering at Nyeeruna's impotence.*

*Thus the myth has come down through the ages, but its special interest to ethnologists lies in its adaptation and re-adjustment to the real lives of the groups "owning" the myth.*

*It has been "dramatised as a performance for men only, and is acted as a comedy or satire before every young initiate. The myth is first recited to them with many un-publishable details and every night during their novitiate the "play" is performed. They see the Nyeeruna actor trying to reach Kambugudha and her young sisters and they watch Babba the dingo disgracing Nyeeruna's manhood before the sisters, and see him crawl away in shame and ignominy. No woman can see or take part in the performance but within an enclosure, just about the distance away in which Nyeeruna and Kambugudha and her sisters "sit down" in the sky, a bush enclosure is made before the play begins and within this enclosure women and girls are hidden and raided at will by all the performers, including Nyeeruna. The women represent Kambugudha and her young sisters and the young novices are taught that they can raid young women at will when they have become men. A Nyeeruna is shown throughout in the drama as a "shocking example" to all men.*

*During the performance songs are sung by the groups owning the special myth, the songs being accompanied by the beating of short heavy clubs on a prepared sand mound (mankind's first "drum") the drum beating and singing being quick and loud or slow and soft as the drama proceeds, the frequent "raiding" of Kambugudha and her sisters being hailed with triumphant drumming and singing.*

*This performance usually begins when the young boys are considered ready for initiation, and at a period when Nyeeruna is absent from the night sky, and it may last until Nyeeruna becomes visible again.*

*Night or day every native of the group owning the myth can point out the exact position of Nyeeruna and the other stars and constellations. The young initiates are thoroughly taught Nyeeruna's story, which they must never reveal to women. The moral of the story is meticulously explained by the brothers or guardians of each young novice.*

*The boys must look upon all women as their slaves, to do their will at all times and in all places, to "fetch and carry" for them throughout their lives.*

*A certain ruthless and savage power is thus instilled into the young novices as they fully grasp - through a wearisome reiteration the acted story of the constellation, and see it turned topsy-turvy in meaning and application, and when they realize their appalling power over all their women-kind and think of Kambugudha's successful defiance of Nyeeruna's advances, whatever cruelty is inherent in them in given full bent.*

*The myth and performance (both grossly phallic) cover a wide area of Central Australia and the western border, south towards the Great Plain's northern edge and east and Southeast towards the Diamantina, Cooper and other rivers.*

*There is a religious instinct, though in a debased form, in this myth, as their only religious sentiments centre round phallicism. Totems, legends, initiation, all rites and ceremonies are representations of phallic worship.*

**About the Authors**

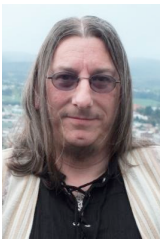

Trevor Leaman is a PhD candidate the Nura Gili Indigenous Programs Unit, University of New South Wales in Sydney, Australia. He is researching the astronomical traditions of the Wiradjuri people of central NSW under the supervision of Dr Duane Hamacher. He earned degrees and diplomas in biology, forestry, engineering, and astronomy, and works as an astronomy educator at Sydney Observatory.

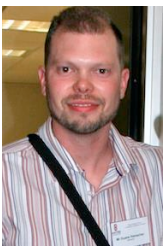

Dr Duane Hamacher is a Lecturer and ARC Discovery Early Career Researcher at Nura Gili and is founder and Chair of the Australian Society for Indigenous Astronomy. His research and teaching focuses on cultural astronomy and geomythology, with a focus on Australia and Oceania. He earned graduate degrees in astrophysics and Indigenous studies and works as a consultant curator at Sydney Observatory.